\documentclass[12pt]{article}

\usepackage[ruled,vlined]{algorithm2e}
\usepackage{graphicx}
\usepackage{bm}
\usepackage{pdflscape}
\usepackage{hyperref}
\usepackage{changes}

\begin{document}

\title{Fast unfolding of communities in large networks: 15 years later}

\author{Vincent Blondel\textsuperscript{1}, Jean-Loup Guillaume\textsuperscript{2} and Renaud Lambiotte\textsuperscript{3}}

\date{\footnotesize{
\textsuperscript{\textbf{1}}UCLouvain, Belgium, vincent.blondel@uclouvain.be\\
\textsuperscript{\textbf{2}}L3i, La Rochelle Université, France, jean-loup.guillaume@univ-lr.fr\\
\textsuperscript{\textbf{3}}Turing Institute, University of Oxford, London, UK, renaud.lambiotte@maths.ox.ac.uk}}

\maketitle

\begin{abstract}
The Louvain method was proposed 15 years ago as a heuristic method for the fast detection of communities in large networks. During this period, it has emerged as one of the most popular methods for community detection, the task of partitioning vertices of a network into dense groups, usually called communities or clusters. Here, after a short introduction to the method, we give an overview of the different generalizations and modifications that have been proposed in the literature, and also survey the  quality functions, beyond modularity, for which it has been implemented. 
\end{abstract}

%
\vspace{2pc}

\section{Introduction}

Networks provide a powerful language to model and analyse interacting systems, in a broad range of disciplines~\cite{newman2018networks}. Facebook and food-webs, the brain and the job market are all systems where connectivity is critical, and where elements -- here users, species, neurons and workers -- are represented by vertices, while pairwise interactions -- here friendship, predation, synchronization and job mobility -- are represented by edges. Over the last two decades, several tools have been developed to extract information from the myriads of interactions composing large networks. Among those, community detection has played plays a central role. Community detection aims at uncovering the community structure of a network, that is its organization into groups where vertices in the same group are strongly connected with each other, and weakly connected with vertices outside the group~\cite{fortunato2010community,fortunato202220}. 
Community detection is usually considered as an unsupervised clustering problem where the purpose is to find the best -- according to a given definition -- division into groups together with the most appropriate number of groups. Community detection is a rich and challenging problem, that has been considered from a variety of perspectives and principles~\cite{schaub2017many} and finds many applications, from network visualization~\cite{garza2019community} to vertex classification~\cite{devooght2014random}. In its most popular form, the communities are assumed to form a partition of the network, so that each vertex belongs to one single community. Partitioning methods have the advantage that their outcome is easier to interpret than that of overlapping methods, but also to involve less parameters to estimate and to have sound mathematical foundations. Even within the family of partitioning methods, several complementary approaches have been developed, leading to different functions to measure the quality of a partition defined in terms of e.g. the density of links inside communities~\cite{newman2004finding}, the number of links between communities~\cite{von2007tutorial}, the confinement of random walkers inside communities~\cite{rosvall2008maps,lambiotte2014random} or its likelihood of having generated the observed network~\cite{peixoto2013parsimonious}.

A common challenge across methods is the need for efficient algorithms to optimize their respective quality function, that is finding the partition of the network with an optimal value. This problem is not new and finds its root in graph partitioning, where different algorithms have been proposed since the 1970s. It is more recently, arguably since the seminal works of Newman~\cite{newman2004finding} that the research community started focusing on the related problem of community detection. It rapidly gained popularity due to the increasing access to large network data whose analysis required efficient clustering methods. The most popular quality function is the so-called Newman-Girvan modularity, and community detection is often considered - in practice - synonymous to modularity optimization, despite its limitations and the availability of alternative approaches. Until 2007, the most efficient methods for community detection were either spectral, i.e. defining modules based on dominant eigenvectors of matrices associated to the graph, or greedy, with the early works of Clauset et al.~\cite{clauset2004finding} and  Wakita and Tsurumi~\cite{wakita2007finding}.
However, the former are by construction limited to networks of the order of $10^4$ vertices, orders of magnitude smaller than large-scale social and information networks, whereas the proposed greedy methods did not perform sufficiently well in terms of accuracy, due to the ruggedness of the optimization landscape. 

\begin{figure}[ht]
\centering
\includegraphics[width=0.9\textwidth]{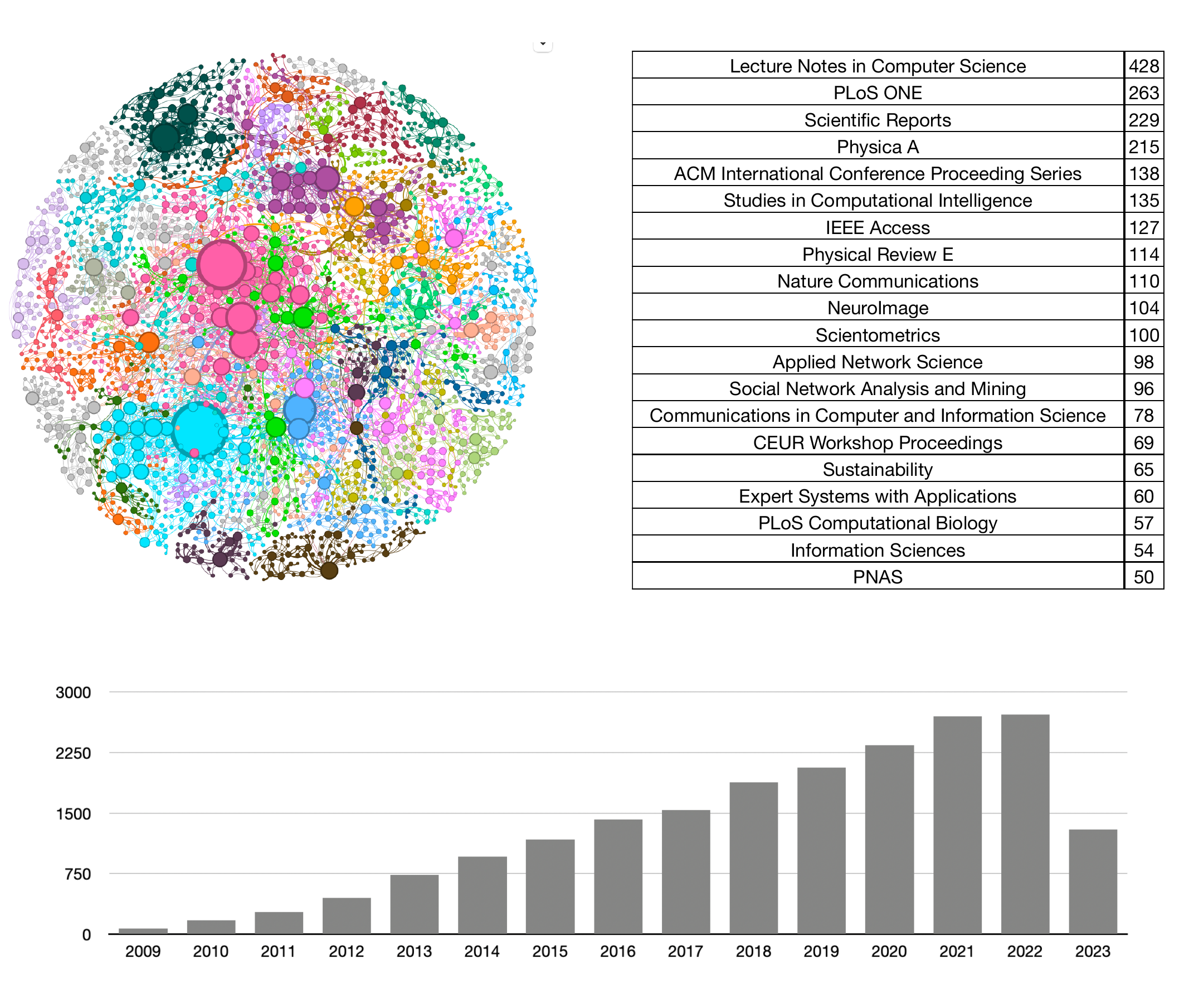}
\caption{
(Top left) Collaboration of researchers having cited the Louvain method. All articles citing~\cite{blondel2008fast} have been downloaded from Scopus on the 6th of April 2023. A bipartite network has been constructed relating author IDs, as identified by Scopus, and papers. The collaboration network was then constructed by first projecting the network according to~\cite{newman2001scientific}, removing edges with a weight smaller than $0.2$ and keeping the largest connected component. The resulting network is composed of 2490 vertices. Running the Louvain method identifies 39 communities, with a modularity of 0.926. The size of the vertices is proportional to their weighted degree. (Top right) Originally published in a Physics journal, the Louvain method has found applications in a variety of scientific domains, as shown by the list of the top 20 journals citing~\cite{blondel2008fast}. Further down in the list, one can find journals as diverse as Poetics (9 articles), Journal of Transport Geography (8 articles) and Nature Immunology (5 articles). (Bottom) Histogram of the total number of citations per year according to Google Scholar (20,054 citations in total on July 6 2023).}
\label{fig:collaborations}
\end{figure}

On March 4 2008, we posted on the arXiv an article entitled ``Fast unfolding of communities in large networks " proposing a greedy method for modularity optimization, and made available an efficient C++ code. A workshop was organized a few days later, on  March 8, entitled ``Detection and visualization of communities in large complex networks", where we invited the leading researchers in the field to our research group in Louvain-la-Neuve. After a desk rejection from Nature and a swift rejection from PNAS, the paper was submitted to JSTAT, accepted on September 3 and published on October 9~\cite{blondel2008fast}. 15 years later, the paper has been cited just over 20,000 times on \href{https://scholar.google.com/scholar?cites=1787079529654689362}{Google Scholar}, this is one new citation every 6 hours for more than 15 years. The method had no name in the original article, and was for a time named BGLL after the authors' initials. The term "Louvain method" appears for the first time in an article published in September 2009~\cite{lambiotte2009communities}. The Louvain method, implemented in standard libraries for graph mining (see Table~\ref{table:omplementations}), is at the core of the network science ecosystem, with applications in several fields of science (see Figure~\ref{fig:collaborations}).  The purpose of this perspective paper is take a step back and to consider the legacy of the paper, by giving an overview of the research surrounding it, from improvements of the original method to its applications to other quality functions.

\begin{landscape}
\begin{table}[ht]
\begin{center}
\begin{tabular}{||c | c | c||} 

    \multicolumn{3}{c}{Libraries} \\
    \hline
    Language & Link & Objective function \\ [0.5ex] 
    \hline\hline
    Python & \url{https://networkx.org/} & mutiresolution $Q$, directed  \\ 
    \hline
    Python & \url{https://networkit.github.io/} & mutiresolution $Q$, parallel  \\ 
    \hline
    Python & \url{https://github.com/vtraag/leidenalg} & several functions, Leiden \\
    \hline
    Julia  & \url{https://github.com/afternone/CommunityDetection.jl} & several functions \\
    \hline
    Javascript & \url{https://graphology.github.io/} & mutiresolution $Q$ \\
    \hline
    Matlab & \url{https://github.com/GenLouvain/GenLouvain} & several functions, multilayer \\
    \hline
    Java   & \url{https://github.com/jcnguyen/research-2015} & several functions\\
    \hline
    
    Multiple & \url{https://igraph.org/} & mutiresolution $Q$  \\ 
    \hline 
    \multicolumn{3}{c}{} \\
\end{tabular}

\begin{tabular}{||c | c | c||} 
    \multicolumn{3}{c}{Softwares} \\
    \hline
    Name & Objective & Link \\ [0.5ex] 
    \hline
    Cytoscape (plugins) & Analysis and visualisation & \url{https://cytoscape.org/}  \\
    \hline
    Gephi & Analysis and visualisation  & \url{https://gephi.org/} \\
    \hline 
    Tulip & Analysis and visualisation  & \url{https://tulip.labri.fr/site/} \\
    \hline
    Pajek & Analysis and visualisation  & \url{http://mrvar.fdv.uni-lj.si/pajek/} \\
    \hline
    SAS Viya & Analytics & \url{https://www.sas.com/en_us/home.html} \\
    \hline
    Neo4j & Database & \url{https://neo4j.com/fr/} \\
    \hline
    Memgraph & Database &  \url{https://memgraph.com/} \\
    \hline

\end{tabular}

\end{center}
\caption{List of implementations of the Louvain method.}
\label{table:omplementations}
\end{table}
\end{landscape}

\section{Basics on networks and modularity}

A network is represented mathematically by a graph $\mathcal{G}(V,E)$, made of a set of vertices $V$ of cardinality $n \equiv |V|$, and a set of edges $E = \{ \{i,j\}\; |\; i,j \in V\}$ of cardinality $m \equiv |E|$. Without loss of generality we will identify the vertex set $V$ with the set $\{0,\ldots,n-1\}$. In this introductory section, we will assume for the sake of simplicity that the network is undirected, so that edges do not have a direction, and unweighted, so that the graph only encodes the presence or absence of an edge between two vertices. Note that the concepts described in this section, and in particular expressions for modularity, have been generalized to more general settings, as we discuss in detail in Section~\ref{subsec:improvement_nonsimple}. A useful way to encode and investigate the structure of a graph is its adjacency matrix. The adjacency matrix $\bm{A}$ is an  $n \times n$  matrix encoding the presence or absence of an edge between any pair of vertices in the graph. Its entries $A_{ij}$ are equal to 1 if $i$ is adjacent to $j$, and 0 otherwise. For undirected graphs we have $A_{ij} = A_{ji}$, i.e., the adjacency matrix is symmetric ($\bm A=\bm A^\top$). Several properties of a graph can be readily calculated from the adjacency matrix, including the degree of each vertex $i$ as $k_i = \sum_j A_{ij}$, but above all, this representation makes it possible to exploit concepts and tools from linear algebra. 

Since the seminal works of Watts and Strogatz~\cite{watts1998collective}, it has been recognised that real-world networks exhibit a mix of random and non-random patterns. Important types of structures, found in a wide range of empirical data, include clustering, here understood in terms of the density of triangles, fat-tailed degree distribution, power-law or not~\cite{broido2019scale}, and  the presence of communities in a network~\cite{simon1977organization}, that is the presence of dense clusters that are weakly connected with one another. Community detection is an algorithmic problem aiming at finding the community structure of a network. As such, it can be seen as an unsupervised clustering problem, and some methods actually consist in first embedding the vertices in a space, before applying classical clustering methods, such as $k$-means~\cite{macqueen1967kmeans}. In this section, we describe one of the most popular methods for community detection, despite some limitations described below, based on the notion of Newman-Girvan modularity~\cite{newman2004finding}, modularity in short.

Modularity is a measure of the quality of the partition of network into groups. Let us first consider a group of vertices defined by a set $\alpha$. The intuitive idea behind modularity is to compare the number of links inside $\alpha$ with an expectation of this number under a random null model. The choice of null model is in principle open and should depend on the mechanisms that constrain the formation of edges~\cite{expert2011uncovering}. The most popular choice is, by far, the soft configuration or Chung-Lu model~\cite{chung2002connected}, a random graph model fixing, on expectation, the degrees of each vertex. For this choice, the expected number of edges between two vertices $i$ and $j$ is given by $\frac{k_ik_j}{2m}$, and the contribution of community $\alpha$ to modularity is given by
\begin{equation}
    \sum_{i, j \in  \alpha}
\left(A_{ij}-\frac{k_ik_j}{2m}\right).
\label{eq:compareQ}
\end{equation}
Note that each term in the sum may either be positive or negative depending on the presence or absence of a link between two vertices. Intuitively, a good, dense community will be one with a high value for this expression. 
Note also that the null model penalises more pairs of vertices with a higher degree, which tends to favor configurations where high degree vertices are placed in different communities~\cite{lambiotte2014random}.

The Newman-Girvan modularity does not focus on one single community, but on the partition of the vertices into $C$ communities $\mathcal{P} = \{\alpha_1,\alpha_2,\ldots,\alpha_C\}$. The quality of a partition is obtained from the sum of Eq.(\ref{eq:compareQ}) over the communities, leading to the expression

\begin{equation}
Q = \frac{1}{2m}\sum_{\alpha \in \mathcal{P}}
\sum_{i, j \in  \alpha}
\left(A_{ij}-\frac{k_ik_j}{2m}\right),
\end{equation}
where the additional prefactor  $1/(2m)$ ensures that modularity is in the interval $[-1,1]$. Note that modularity can also be interpreted in terms of the assortativity coefficient, a correlation measure of vertex attributes, here the community labels,  at the endpoints of edges in the network~\cite{peel2018multiscale}, in terms of a covariance defined on the graph~\cite{devriendt2022variance} and in terms of the clustered covariance matrix from the random walk exploration of the graph~\cite{delvenne2010stability}. Intuitively, modularity  gives a high value to a partition if its communities concentrate an unexpectedly large number of edges inside its communities. In contrast with  classical concepts such as the cut size, modularity allows to compare partitions made of different numbers of communities, and the partition maximising modularity is usually non-trivial, i.e. neither made of one large community or of $n$ singletons. Modularity has become an essential element of several community detection methods. Modularity optimization methods aim at optimising modularity, that is finding the ``best" partition of a network having the highest value of modularity. As modularity optimization is NP-hard~\cite{brandes2007modularity}, several heuristics have  been proposed for modularity optimization. Popular techniques include spectral methods~\cite{newman2013spectral} but large networks require faster, greedy methods proceeding by agglomerating groups of vertices into larger ones ~\cite{Fortunato2010}.

Before going further, note that the contribution of vertex $i$, assigned to a certain modularity $\alpha$, to the modularity of a partition is given by 
\begin{equation}
\label{local_Q}
Q_i = \frac{1}{2m}
\left( \sum_{j \in  \alpha} A_{ij}-\frac{k_i k_\alpha}{2m}\right),
\end{equation}
where $k_\alpha$ is the degree of community $\alpha$, defined as the sum of the degrees of its vertices. This expression shows that the contribution of vertex $i$ requires only local information, its number of neighbors inside the community, its degree and the degrees inside the community.  Locality is essential to design efficient greedy algorithm for modularity optimization.

Modularity, as a quality function, suffers from some limitations, which are inherited by any modularity optimization algorithm. Among those, let us note its tendency to over-fit, finding communities even in the case of random graphs~\cite{guimera2004modularity}, and to have, implicitely,  a characteristic scale making it search for communities with a size that might not be compatible with the underlying structure~\cite{ghasemian2019evaluating}. Generalizations of modularity have been proposed to alleviate these limitations, for instance by incorporating a resolution parameter allowing to tune the characteristic size of the communities~\cite{reichardt2006statistical,delvenne2010stability,lambiotte2009laplacian}.

\section{The Louvain method}

The Louvain method is a greedy optimization heuristic originally proposed to optimize modularity, but its flexibility allows to optimize other quality functions as well, as described below. The approach consists in two phases that are iteratively repeated, until a  maximum of modularity is obtained. The first phase consists in moving vertices from their community to that of their neighbors until a local maximum is reached. The second phase then consists in forming a new graph whose vertices are obtained by aggregating the vertices attached to the same community in the previous phase. A combination of the two phases is called a "pass". Starting from a network composed of $n$ vertices, the number of vertices decreases at each pass, until no improvement of modularity can be obtained.

{\bf First phase: vertex mover (VM).}
The first phase begins with an undirected weighted graph having $C$ vertices to which an index between $0$ and $C-1$ has been randomly assigned. $C$ is the number of vertices in the first pass, and the number of communities found in previous pass otherwise. One starts by placing each vertex into its own community, that is we start with a partition made of $C$ communities. We then consider the first vertex, i.e. with index $0$, and evaluate the change of modularity by removing $0$ from its community and placing it in the community of each of its neighbors. Importantly for the speed of the method, this step only requires local information -- See Eq.(\ref{local_Q}) --, and its number of operations is proportional to the degree of the vertex. Vertex $0$ is then assigned to the community where the increase is maximum, if this maximum increase is positive. Otherwise, vertex $0$ is left in its original community -- See Algorithm~\ref{algo:singlevertexmover} (all algorithms are adapted from~\cite{campigotto2014generalized}). This step is repeated sequentially to all the vertices based on their index. When reaching vertex $C-1$, the process restarts at vertex $0$, and the iteration continues until no vertex is moved during a complete iteration over the $C$ vertices -- See Algorithm~\ref{algo:vertexmoveriteration}. The phase ends in a local maximum, where greedy move of the vertices between community do not allow for an increase of the modularity. At this moment, a new partition of the vertices into $C^{'}$ communities has been produced. If $C^{'} \neq C$, one proceeds to the next phase, and update the value of the number of communities. Else, the algorithm is finished and returns the partition as its estimate for optimum of modularity.

\begin{figure}[ht]
\centering
\includegraphics[width=0.8\textwidth]{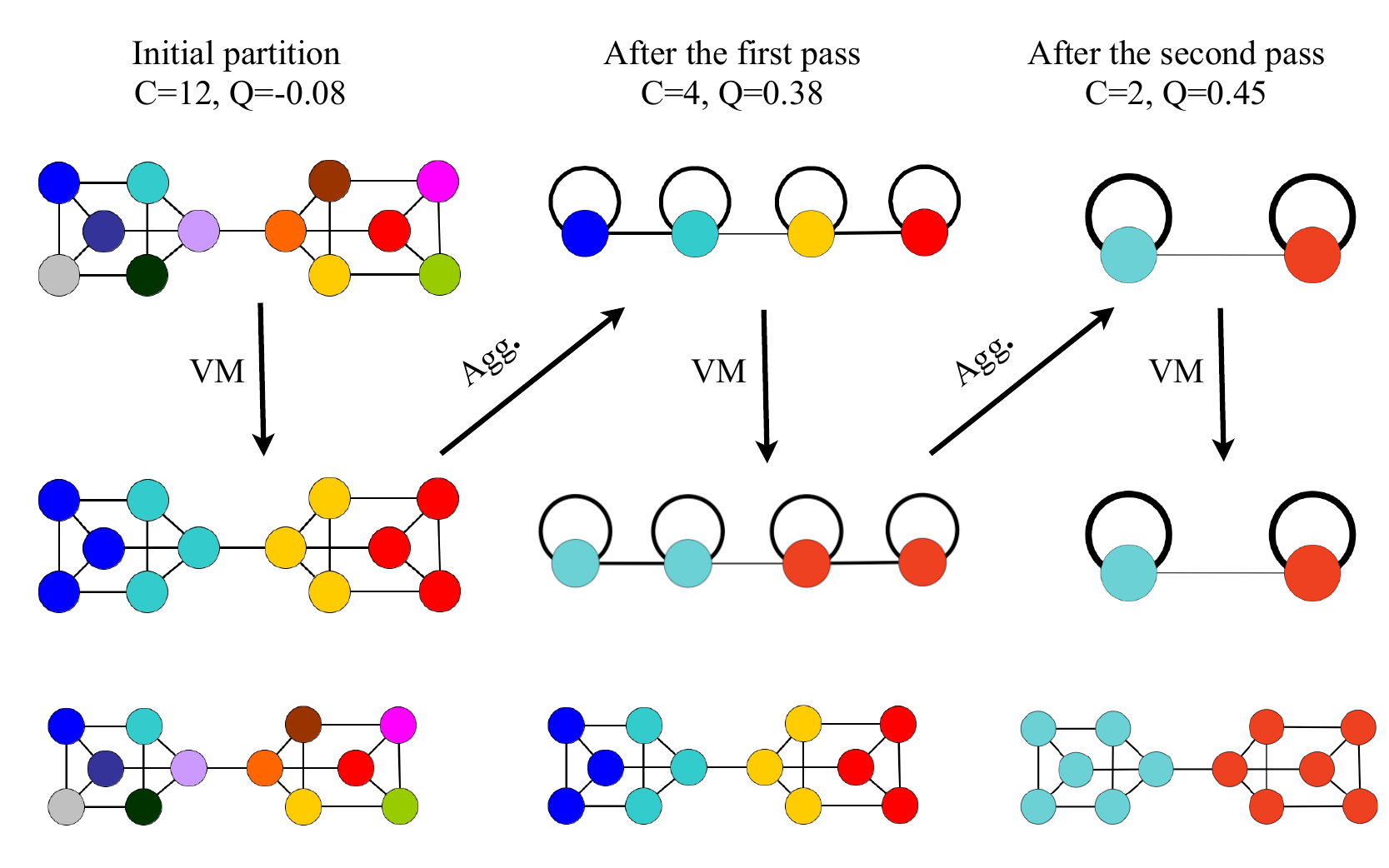}
\caption{
Each pass of the  Louvain method is made of two phases. The first phase (Vertex Mover) consists in sequentially moving vertices to neighboring communities leading to a maximum increase of modularity. The second phase aggregates vertices and constructs a meta-graph whose vertices are the communities found after the first phase. The two phases are repeated until no improvement of modularity is observed. In this illustration, the algorithm converges after two passes, uncovering an optimal partition made of 2 communities, and a value of $Q=0.45$. In this example, it is easy to show that aggregating further the two communities into one big community would result in a decline of modularity, as $Q=0$ in that case.
Figure adapted from~\cite{aynaud2013multilevel}.}
\label{fig:explication}
\end{figure}

{\bf Second phase: aggregation.}
The second phase consists in building a new weighted graph whose vertices are the $C$ communities found during the first phase. The weight of the link between two communities is given by the sum of the weights of the links between the vertices of these two communities. 
Moreover, links between the vertices of a same community lead to self-loops in the aggregated network. 
This operation is performed to ensure that the modularity of a partition in the aggregated graph has the same value as that of the same partition on the original graph. 
Once the second phase is finished, the first phase is reapplied on the aggregated network.

The output of the algorithm is the partition uncovered in the last pass, as it is the best partition found in terms of modularity -- See Figure~\ref{fig:explication} and Algorithm~\ref{algo:louvain}. The method can also provide intermediary steps, that is the set of partitions of each pass, which provides a hierarchical representation of the network, each community at a pass being composed of a combination of combinations at the previous pass. Since its inception, the method has consistantly been shown to provide a good balance between accuracy and speed of execution, and to have a complexity close to linear in benchmarks and empirical networks. The speed of the method is ensured by the locality of each operation, as vertices only move to neighboring communities. Its accuracy arises from its flexibility, as vertices may be removed from their community and re-assigned to others in the process, and its multi-scale nature, as the community exploration is first done locally, and then over longer distance as the vertices are aggregated. Combined with its simplicity, these strengths have made the Louvain method a popular choice to optimize not only modularity, but also other quality functions, including the Map Equation~\cite{rosvall2008maps} and Markov stability~\cite{lambiotte2009laplacian}, as we discuss more in detail below. When applied on modularity, as is usually the case, it is important to remember that the Louvain method inherits the limitations of modularity, including its resolution limit.

\begin{algorithm}[ht]
  \caption{SingleVertexMover}
  \label{algo:singlevertexmover}
  
  \DontPrintSemicolon
  
  \SetKwInput{Require}{Require}
  \SetKwInput{Ensure}{Ensure}
  \SetKwInput{Data}{Local}
  
  \Require{$G=(V,E,w)$ a weighted graph}
  \Require{$\mathcal{P}$ a partition of $V$ ($\mathcal{P}[i] = $ community of vertex $i$)}
  \Ensure{$\Delta Q$, the gain of modularity}
  \BlankLine  
  \Begin{
    $c_{old} \leftarrow \mathcal{P}[i]$\;
    REMOVE$(i,c_{old})$\;
    $\mathcal{C} \leftarrow \{\mathcal{P}[j] \mid (i,j)\in E\} \cup \{c_{old}\}$\;
    $c_{new} \leftarrow \arg\max_{c \in \mathcal{C}}\{$GAIN$(i, c)\}$\;
    \If{$GAIN(i, c_{new}) - LOSS(i, c_{old}) > 0$} {
        INSERT$(i,c_{new})$\;
        return $GAIN(i, c_{new}) - LOSS(i, c_{old})$\;
    }
    \Else{
        INSERT$(i,c_{old})$\;
        return 0\;
    }
  }
\end{algorithm}

\begin{algorithm}[ht]
  \caption{VertexMoverIteration}
  \label{algo:vertexmoveriteration}
  
  \DontPrintSemicolon
  
  \SetKwInput{Require}{Require}
  \SetKwInput{Ensure}{Ensure}
  \SetKwInput{Data}{Local}
  \SetKwRepeat{Do}{do}{while}%
  
  \Require{$G=(V,E,w)$ a weighted graph}
  \Require{$\epsilon$ a stopping criterion}
  \Ensure{a partition $\mathcal{P}$ of $V$}
  
  \BlankLine
  
  \Begin{
      INIT$(\mathcal{P})$\;
      \Do{$increase > \epsilon$}{
        $increase \leftarrow 0$\;
        \ForAll{\textup{vertices} $i$}{
          $increase \leftarrow increase + $SingleVertexMover(i)\;
        }
      }
      return $\mathcal{P}$\;
    }
\end{algorithm}

\begin{algorithm}[ht]
    \caption{Louvain algorithm}
    \label{algo:louvain}
    
    \DontPrintSemicolon
    
    \SetKwInput{Require}{Require}
    \SetKwInput{Ensure}{Ensure}
    
    \Require{$G=(V,E,w)$ a weighted graph}
    \Ensure{a partition $\mathcal{P}$ of $V$}
    
    \BlankLine
    
    \Begin{
        \Repeat{\upshape no improvement is possible}{
          $\mathcal{P} \leftarrow $VertexMoverIteration$(G)$\;
          $G \leftarrow $PartitionToGraph$(\mathcal{P},G)$\;
        }
      }
      
    \end{algorithm}

\section{Louvain++, a selection of enhancements}

Because of its effectiveness and simplicity, the Louvain method has been extensively studied, and numerous modifications have been proposed, whether to improve results, further increase calculation speed or generalize its scope of use. We present these modifications in four different directions, each detailed in its respective subsection:
\begin{itemize}
\item the modification of the algorithm's core, including its initialization, stopping criterion, vertex mover, etc;
\item the use of quality functions other than modularity, functions which may be similar to modularity but with other parameters, or functions which may be completely different;
\item generalizations of Louvain to deal with more than simple graphs. These may include weighted, oriented, spatially constrained, temporal networks, etc;
\item and finally the use of parallelism to speed up computation time.
\end{itemize}

\subsection{Modification of the algorithm}
\label{subsec:improvement_algorithm}

The original article and an extended version of it~\cite{aynaud2013multilevel} (published in French in 2011 and in English in 2013) already described several improvements that have been proposed independently and studied in greater details by other authors:
\begin{itemize}
    \item a partial optimization that consists in stopping the optimization when the modularity gain is below a given threshold (both for iterations or aggregation). This very simple optimization was already implemented in the initial version of Louvain;
    \item removing leaf vertices, which consists of deleting the leaves since they can only be placed in the community of their single neighbor;
    \item a VM that consists in not considering all vertices at each iteration;
    \item returning to lower levels to move again vertices;
    \item a simulated annealing-inspired VM allowing for non-optimal moves.
\end{itemize}

In the following, we describe modifications proposed in the different steps of the algorithm, that is its initialization, the vertex mover step and the aggregation step.

{\bf Initialization}:  Several types of optimization have  targeted the initialization phase. The first one is to reduce the size of the graph before computation by removing a number of vertices that have little or no impact on the results. Based on the observation that a leaf (any vertex with a single neighbor) will inevitably be placed in its neighbor's community, several authors propose deleting these leaves~\cite{aynaud2013multilevel, lui2021study} or, more generally, all tree structures, i.e. the initial leaves and all those that will appear when deleting leaves~\cite{zhang2021improved}. The resulting graph corresponds to the 2-core of the original graph and the leaves are added back once the communities are found. A generalization has been proposed in~\cite{peng2014accelerating} where a k-core decomposition is performed and communities are only computed on the k-core. All the removed vertices are then added back and assigned to a community. The authors propose to use label propagation or any modularity maximization algorithm for this task. 

Other approaches have modified the initial partition by not starting with communities containing a single vertex. The authors of~\cite{abbas2020improving} identify cliques containing at least 3 vertices that serve as initial communities. However, the way cliques are identified and the management of overlaps between these cliques are not made explicit.

{\bf Vertex mover:} In the Louvain method, the best choice is always preferred during the vertex mover operation. Based on the principle of simulated annealing, it was proposed to choose a non optimal move, even with a negative gain, with a low probability rather than the optimal one~\cite{zhou2012community}.

Several articles propose to avoid considering all vertices at each iteration. The simplest version is defined in~\cite{ryu2016quick}. If a vertex does not move for several iterations (with a fixed parameter) then it will not be considered anymore. In~\cite{zhang2021improved}, the authors show that if two communities $A$ and $B$ have not changed in the last iteration, then no vertex can move from $A$ to $B$ or from $B$ to $A$ in the current iteration. A similar idea is used in~\cite{ozaki2016simpleacceleration} where the author identify vertices that are more likely to move in the next iteration and only consider these vertices. 
They propose four situations, the simplest of which considers that if a vertex $i$ has just moved from the community $A$ to the community $B$, then all $i$'s neighbors who are not in $B$ are likely to join $i$ at the next iteration. The other three situations are a little more complex and are not used in practice. Limiting oneself to considering only a few vertices at the next iteration could lead to iterations being stopped prematurely and thus to a loss of quality, but in practice the impact is very slight.

In a different fashion, authors of~\cite{lui2021study} use a concept of seed vertices which are vertices more central than the average (for a given centrality measure defined in the paper). Non-seed vertices are then moved in priority to the neighboring community that contains a seed-vertex and maximizes the (positive) gain of modularity. If there is no such neighboring community or if no gain is positive, then the classical vertex mover is used.

In~\cite{traag2015faster}, instead of considering all neighbor communities and pick the best, one neighbor is picked at random and only its community is considered -- other random selection procedure have also been considered, e.g. a community is selected proportionally to its size. Even though random selection will not always make the best choice, if a vertex has many neighbors in a given community then this community is more likely to be picked. The author notes that this operation accelerates convergence. This principle has been reused in the Leiden method~\cite{traag2019leiden} with other elements.

In most implementations of Louvain, the order in which vertices are considered is random. Although this can be seen as a problem, it does allow us to explore more possible solutions. If the aim is to obtain an identical partition, any fixed order can be used. In~\cite{aldabobo2022improved} the authors treat the vertices by decreasing centrality. This choice appears to improve results slightly, but above all, it speeds up the algorithm.

Other improvements have been proposed to simplify VM in a distributed computing context. We describe them in Section~\ref{subsec:improvement_distributed} 

{\bf Aggregation and refinement:}
A key element of the Louvain method is that the vertex mover operation is not greedy since any given vertex can be moved several times until convergence is reached. As this step is not irreversible, it is less likely to be penalized by poor initial choices. However, the aggregation phases are irreversible, and the original method does not allow to go back once the graph has been aggregated. For this reason, a number of papers have proposed to challenge this principle and allow to move lower-level vertices even after aggregation. This procedure is called partition refinement.

In the simpler versions~\cite{du2016common}, a single refinement of the partition is performed on the first level, i.e., the original graph where all vertices are subject to the vertex mover procedure until convergence.
In~\cite{gach2014improving, rotta2011multilevel}, a multi-level refinement is used where, starting from the top level of the partition, the graph is unaggregated once and vertices are allowed to move. This process is repeated moving down the hierarchy until they are back on the initial graph.
A different refinement is proposed in~\cite{waltman2013smart,rosvall2011multilevel} where the subgraph induced by each community is partitionned using one level or the complete Louvain method. All these parts of each communities are used to build the aggregated network. A more complex scheme is used in the Leiden method~\cite{traag2019leiden} where the vertex mover during refinement is limited and a move is accepted if it increases the modularity even though it is not the best move.
Finally, in~\cite{yao2023constrained}, if small (less than $4$ vertices) or weak (that score low using their quality function) communities are found, these communities are broken into individual vertices and a VM step is applied once.

The Louvain method and its evolutions do not impose a priori constraints on the number or size of communities. If the number of communities is known to be $k$, then~\cite{darst2014improving} propose to stop the aggregation in Louvain at the highest level that contains more than $k$ communities (i.e. the next aggregation would result in fewer than $k$ communities). They then use a greedy method to merge the communities that maximize the modularity gain to obtain exactly $k$ communities. Finally, VM iterations are done until convergence.

\subsection{Modification of the quality function}
\label{subsec:improvement_function}

Despite the shortcomings of modularity, several quality functions have been inspired by it and used as a replacement in the Louvain method. For those, the algorithm remains unchanged after adapting the expression for the modularity change during the VM step. These quality functions mostly consider a positive contribution for intra-community edges and a negative contribution for inter-community edges or in the absence of edges.

\begin{itemize}

\item Several variations of modularity are using a resolution parameter, and can be shown to be equivalent, up to a trivial map, to the so-called Potts modularity ~\cite{reichardt2006statistical}
\begin{equation}
Q_\gamma = \frac{1}{2m}\sum_{\alpha \in \mathcal{P}}
\sum_{i, j \in  \alpha}
\left(A_{ij}- \gamma \frac{k_ik_j}{2m}\right),
\end{equation}
where the resolution parameter $\gamma$ allows to give more or less importance to the negative term, and hence to modulate the size of the communities. This expression has the same optimum as the so-called Markov stability, where the quality of a partition is estimated by its tendency to capture random walkers for long times inside communities
\cite{delvenne2010stability,lambiotte2009laplacian} and where time plays the role of the resolution parameter. Most implementations of Louvain have been generalized for one (or both) of these equivalent variations.

\item The balanced modularity~\cite{cespedes2013comparing, campigotto2014generalized} uses a symmetric version that also considers the absence of links

\[
\frac{1}{2m} \sum_{i,j\in V} \left[ \left( a_{ij} - \frac{k_ik_j}{2m} \right) x_{ij} - \left( \bar{a}_{ij} - \frac{(n-k_i).(n-k_j)}{n^2-2m} \right) \bar{x}_{ij} \right].
\]

\item The authors of~\cite{zhou2012community} use a function that considers both edges within communities and edges between communities. The linear version of their function is

\[
\frac{1}{2m} \sum_{i,j\in V} \left[ \left( a_{ij} - \frac{k_ik_j}{2m} \right) x_{ij} - \beta \left( a_{ij} - \frac{k_ik_j}{2m} \right)^\alpha \bar{x}_{ij} \right],
\]
where $\alpha$ and $\beta$ are resolution parameters that can be tuned to adjust the size of communities.

\item In~\cite{chaudhary2019community}, the authors propose Jaccard Cosine Share Measure (JCSM). For two vertices $i$ and $j$, the main term combines Jaccard (size of the intersection of neighborhoods of $i$ and $j$ divided by the size of their union), Cosine (normalized dot product of rows $i$ and $j$ of the adjacency matrix) and classical modularity ($a_{ij} - k_ik_j/2m$).

\item Other functions mostly based on the number of connections within or between communities have been proposed. For instance~\cite{yao2023constrained} propose a new function defined as the sum for each community of 
\[
F2(C_i) = \frac{[d_{in}(C_i)]^2}{[d_{in}(C_i)+d_{out}(C_i)]^2}.
\]

\item Based on other principles than modularity, the map equation introduced in~\cite{rosvall2008maps} uses the length of description of random walks in a network. The equation uses a two levels description to encode both the modules and the vertices. vertices from different modules can use the same code (the authors use the analogy of street names, which may be similar in different cities). The optimization of the map equation follows closely the steps of the Louvain method. Note that the map equation has also been generalized to uncover a hierarchy of communities~\cite{rosvall2011multilevel}.
\end{itemize}

More generally, works have explored the conditions that need to be satisfied by a quality function so that it can be optimized by the Louvain method. In that direction, 
the authors of~\cite{campigotto2014generalized} have used relational coding to describe several quality function using the same formalism. A quality function $F$ is said to be linear if $F(X, A) = \sum_{i,j\in V} \phi(a_{ij})x_{ij}+K$, where $a_{ij}=1$ if $i$ and $j$ are connected ($0$ otherwise), $x_{ij}=1$ if $i$ and $j$ are in the same community ($0$ otherwise), $\phi$ and $K$ are respectively a function and a constant that only depends on the original data. For example,  modularity can be expressed as $Q = \sum_{i,j\in V} \left( a_{ij} - k_ik_j/2m \right) x_{ij}$. If a quality function is linear then the gain obtained using VM can be computed locally, henceforth without loss of efficiency. The authors also study separable quality function, i.e. that can be written as $F(X, A) = \sum_{i,j\in V} \phi(a_{ij})\psi(x_{ij})+K$, where $\psi$ only depends on $X$. If the function $\psi$ only depends on the community of interest then the corresponding quality function can also be used efficiently in Louvain. The authors study five non-linear but separable function and clarify why only some can be optimized efficiently. A different formulation is given in~\cite{schaub2019multiscale}. The Louvain method can be used to optimize any quality function of the form $trace\ H^\top[F-ab^\top]H$, where $H$ is the partition indicator matrix, $F$ is a general matrix derived from the network, and $a$ and $b$ are two $n$ dimensional vectors.

\subsection{Louvain for non-simple graphs}
\label{subsec:improvement_nonsimple}

Modularity has originally been defined for undirected, unweighted networks. Since then, several generalizations have been proposed to allow for more complex graph structures. As compared to the expression Eq.(\ref{local_Q}), the first term often remains unchanged, to count the total number, or total weight, of edges inside communities, but the second term is modified to take into account the additional constrains induced by the graph structure. For instance, in the case of directed networks, the null model is now chosen according to the directed configuration model, to properly estimate the expected number of  edges between two vertices, given their respective in-degree and  out-degree~\cite{leicht2008community}. The Louvain method can then be directly generalized with an adapted modularity gain during VM~\cite{li2018improved}. 
Other examples include the optimization of modularity for spatial networks where, again, the null model of the quality function can be modified in order to account for spatial constraints~\cite{expert2011uncovering}. Alternative approaches keep standard expressions for modularity but modify instead the Louvain method to impose a spatial contiguity of the clusters in the uncovered partition~\cite{wang2021network}.

Community detection is also an active field of study for temporal networks~\cite{cazabet2019challenges,masuda2016guide}. Compared to its static counterpart, communities are now dynamical objects that may encounter different types of change, such as growth or splitting, and algorithms have the additional task to characterize these events. Two different types of approach build on community optimization. First, communities can be uncovered in different snapshots. In that case, some constraints on the communities in adjacent snapshots may  be incorporated in order to ensure their continuity  in time, such as in modularity maximization under estrangement constraint~\cite{kawadia2012sequential}.  
After the communities have been found in each snapshot, additional methods determine, with a certain statistical accuracy, when and how a reorganization  takes place~\cite{greene2010tracking}. A second family of methods aims at finding a decomposition of the network in an extended network in which snapshots are concatenated, e.g. by connecting the same vertex in consecutive snapshots to favor continuity of communities across time. Here, instead of searching communities in each time window, the community detection is performed in one single operation, which can be formulated as a modularity maximization, on a larger, multilayer network~\cite{mucha2010community} and again optimized by the Louvain method~\cite{jutla2011generalized}.

Modularity has also been generalized to quantify the quality of the partition of signed networks. Here, the intuition is place together vertices such that positive edges are concentrated within the modules and negative edges between the modules. In its simplest setting, the signed modularity consists in dividing the signed graph into two graphs, one defined by the positive interactions and another one defined by the negative interactions, and to search to maximize the difference of the modularities of the positive and negative graph~\cite{traag2009community}. The Louvain method can be directly generalized for this quality function, and the only step that needs to be considered with care is VM. Consider a vertex $i$. One expects to find other vertices of the community of $i$ in the neighborhood of $i$ only for the positive edges but its negative neighbors should instead be placed in other communities. Ways to address this complication are either to consider all the communities in the network in the VM step - and not only the neighboring ones - , but this involves a massive slow down of the algorithm, or to consider second neighbors, that is neighbors of neighbors, in the case of  negative edges~\cite{John}.

Finally, some works have focused on attributed networks where nodes carry labels. The aim is to create well-connected communities that are also homogeneous with regard to labels, in order to meet both structural and homophilic criteria. Three families of approaches are proposed in~\cite{chunaev2020community}: early fusion where a new graph is built to encode both topology and attributes, simultaneous fusion where the clustering is performed simultaneously on both domains and late fusion where both clusterings are computed independently before being joined. Among those, 
SAC2 \cite{dang2012community} is an early fusion method where each node retains only its k-nearest neighbors based on a weighted combination of topology and similarity between attributes.  Louvain is then used on this construction. Simultaneous fusion generally involves a composite function of modularity for the structural part and a similarity function for the attributes. This is the case in~\cite{citraro2020identifying} using purity for attributes. Relatedly, SAC1 \cite{dang2012community} uses different similarities for attributes, and~\cite{combe2015attributed} uses inertia. In all cases Louvain is used to optimize these functions. A late fusion is proposed  in~\cite{elhadi2012structure} where structural communities are found with Louvain and attributed ones with k-means. The structural communities are only used if they are sufficiently good, otherwise the attribute ones are used. More generally any consensus-based solution can be used to merge several partitions.

\subsection{Distributed/parallel/GPU solutions}
\label{subsec:improvement_distributed}

Many efforts have been made to parallelize the Louvain method, the main challenge being to perform the VM step on several CPUs simultaneously. There are two main paradigms: shared-memory implementations, where processes share the same memory and must therefore synchronize for writing, and distributed-memory implementations, where processors communicate via messages  to maintain a consistent view. The former is limited by the number of physical processing units that can be used, and the latter by the quantity of messages sent.
In most distributed implementations, if not all, the input network is partitioned so that vertices and edges are distributed evenly among the processing units.
Evaluating the correctness of a parallel implementation can be complex. Because of the impact of the order in which vertices are processed, one does not expect to obtain the same results from one run to the next, or when compared with a sequential run.

\noindent{\bf Shared-memory implementations:}
In shared-memory implementations several vertices evaluate simultaneously and independently the best VM. They all base their choice on a structure that may no longer be valid at the time of the decision. The assignment of vertices to communities is protected and can only happen in a sequential way so that no inconsistency can occur but a vertex $i$ may have been assigned between the time a vertex $j$ made its choice and the time it validated it. Hence convergence might be slower or even not attainable. Several approaches have been considered.

\begin{itemize}
   
\item To keep the VM sequential,~\cite{bhowmick2013template} propose an implementation by identifying the loops in the code that can be executed in parallel. This includes initialization of the algorithm, identifying all the neighboring communities of a vertex and finally identifying the neighboring community with the best gain. Community merging can also be partially parallelized.

\item To avoid conflicts that may results in a slower convergence one solution would be to find vertices that do not conflict with each other and parallelize them. While the extra cost is generally considered too high~\cite{shi2021scalable}, some authors propose to identify isolate sets of vertices, i.e. vertices that are not neighbors of each other, and to process them in parallel~\cite{qie2022isolate}. The authors of~\cite{shi2021scalable} propose to neglect this risk of non-convergence but limit the number of iterations. They also compare asynchronous moves with synchronous moves, where all parallely processed vertices are moved simultaneously.

\item The authors of~\cite{fazlali2017adaptative} propose to adapt the number of processes dynamically depending on the number of neighbors of a vertex and available resources.

\item In~\cite{tithi2020prune}, the authors use techniques already presented to avoid considering all vertices at every iteration. In the first iterations they use a pull method, which is the classic method where neighboring communities are recalculated at each iteration for every vertex. In the later iterations there are fewer vertex moves, they switch to a push method where neighboring communities are simply updated (pushed) with each vertex move. This optimization is combined with classic code parallelization, using locks for updates of neighboring communities.

\end{itemize}

\noindent{\bf Distributed memory implementations:} 
In distributed memory implementations, vertices are distributed among the various processes. Each process will therefore store a given number of vertices. A number of problems arise that require algorithmic adaptation of the Louvain method. First, vertices or edges must be distributed evenly between processes to improve overall calculation time. In complex networks, the presence of very large degree vertices (or hubs) makes vertices/edges distribution more complex.
Second, if the ends of an edge are not on the same process, messages will have to be sent to synchronize them and the total number of messages must be kept as low as possible.
Finally, since calculations are made independently, it is necessary to be able to question local choices. Several works have looked at, and addressed, these issues. 

\begin{itemize}
\item In~\cite{cheong2013hierarchical}, the authors use three levels of parallelism: (1) the network is divided in subgraphs and edges between these subgraphs are ignored. Louvain is applied on each subgraph. Then the individual results are joined, reintroducing the missing edges, and Louvain is reapplied. (2) vertices of each subnetwork are considered in batches and vertices of a given batch are processed in parallel. Potential conflicts if two vertices of a batch are neighbors are not considered but the update of vertex's community are still performed sequentially to avoid inconsistency. (3) Modularity gain for neighboring communities is computed in parallel.

\item Processes can store knowledge of the neighbors of the vertices it owns even though such vertices are stored on another device. Such vertices are called "ghost vertices". Similarly, a process can store information about "ghost" communities adjacent to the ones it owns. Ghost vertices and communities are used in~\cite{ghosh2018distributed}. Initially, contiguous block of vertices are distributed trying to balance the number of edges received by each process. At each iteration, some messages are exchanged between processes owning ghost vertices to update the information about community assignment. Graph coarsening is also performed in a distributed way so as to determine new global ids of communities and to exchange partial edge lists of neighbors communities so that each process can have a complete edge list. A similar procedure adapted for GPUs is detailled in~\cite{gawande2022towards}

\item The authors of~\cite{que2015scalable} use two hash tables to store the graph and also propose to identify the vertices that will provide the largest improvement of modularity and to only consider such vertices to accelerate the VM step. They show that the amount of such vertices decays exponentially fast so that only the first few iterations are costly.

\item The initial distribution of vertices can have an impact on the workload but also on the number of ghost vertices and therefore on the quantity of messages to be transmitted. This is particularly the case for hubs so it is worth choosing the initial partition carefully. In~\cite{cheong2013hierarchical}, the authors distribute vertices evenly based on their id. This is studied in more details in~\cite{zeng2015parallel} where the authors show that the degree distribution of such a partition is close to the degree distribution of the whole network, which ensures a fair workload. Then they propose an enhancement that also considers the degree of ghost vertices to improve workload fairness.

\item To account for hubs it is also possible to duplicate them. Each duplicate (called a delegate) retains the neighbors belonging to its partition by default, but this can be modified to ensure a better balance of edges between partitions~\cite{pearce2014faster}. Then, delegates communicate with each other to ensure synchronization. This general principal is used in~\cite{zeng2018scalable} as a preprocessing step for a distributed Louvain implementation. The authors also take delegates into account when calculating the modularity gain.

\item The authors of~\cite{wickramaarachchi2014fast} use PMETIS, a parallel implementation of METIS~\cite{karypis1998fast} to obtain a partition of the network and distribute theses parts. Each one is then processed independently, then aggregation is performed and the rest of the algorithm is executed sequentially.

\item Independent calculations on each process can result in incorrect communities if vertices are incorrectly assigned to a process. This problem is studied in~\cite{bhowmick2022scalable} with the identification of doubtful vertices. The relative commitment for a vertex is calculated by computing the ratio between its internal degree and the maximum internal degree in its community, and a vertex's affinity for a community is calculated by taking into account the relative commitment of the vertex and its neighbors. Vertices with low commitment are considered doubtful and can be moved to another device if they have many connections with vertices on that device.

\end{itemize}

\section{Perspectives}

In this article, we have overviewed the different ways in which the Louvain method has been used, generalised and improved over the last 15 years. As we have shown, a wide variety of works have built on its original design, exploring how to use it for other quality functions than modularity, and adapting its different steps for faster, more accurate and/or distributed solutions. A great deal is known about the strengths and weaknesses of the methods as well as its theoretical foundations, yet it remains an active source of inspiration as well as a standard method to cluster large-scale networks in practice.
Never did we imagine the impact of the method when it was originally released and it is equally difficult for us to foresee now what will become of it in the years to come. The method has accompanied the different trends that have energed in network science,  from multiplex networks to embeddings and temporal networks. For this reason, it is likely that the Louvain method will find applications and generalisations in the increasingly popular field of higher-order networks~\cite{lambiotte2019networks,battiston2020networks,bianconi2021higher,bick2023higher}, but also in continuous generalisations of networks like graphons~\cite{caron2017sparse} and in extensions of graphs to non-real weights~\cite{bottcher2022complex,tian2023structural}. But we are even more sure that we will be surprised, and
we look forward to sitting down in 15 years, discussing together, discovering in awe the many ways the research community has modified our original idea and, who knows, sharing it with you~\cite{blondel2038}.

\section*{Acknowledgments}

Our special thanks go to Etienne Lefebvre, without whom the Louvain method would never have existed. Only one publication as a master student, but what a success!
We also thank John Pougu\'e Biyong for his help to analyse the Scopus data and the many colleagues with whom we have collaborated on this topic over the year, in particular Thomas Aynaud, Mauricio Barahona, Romain Campigotto, Patricia Conde-Céspedes, Jean-Charles Delvenne, Mason Porter, Martin Rosvall and Michael Schaub.
Finally, many thanks to Sébastien Amoury and Shazia Babul for carefully reading this manuscript.

The work of RL was supported by the EPSRC grants EP/V013068/1 and EP/V03474X/1.
The work of JLG was supported by the ANR MITIK project, French National Research Agency (ANR), PRC AAPG2019.

\bibliographystyle{iopart-num}
\bibliography{main.bib}

\end{document}